\documentclass[10pt,conference]{IEEEtran}

\usepackage{cite}
\usepackage{amsmath,amssymb,amsfonts,amsthm}
\usepackage{algorithmic}
\usepackage{graphicx}
\usepackage{textcomp}
\usepackage{xcolor}
\usepackage[hidelinks]{hyperref}
\usepackage[short]{optidef}
\usepackage{mleftright}
\usepackage[capitalize]{cleveref}

\let\origbibitem\bibitem
\renewcommand{\bibitem}[1]{\origbibitem{#1}\looseness=-1\relax}

\newtheorem{lemma}{Lemma}

\theoremstyle{definition} 
\newtheorem{example}{Example}
\newtheorem*{remark}{Remark}

\renewcommand{\S}{\mathcal S}
\DeclareMathOperator{\A}{\mathcal A}

\renewcommand{\L}{\mathcal L}
\newcommand{\R}{\mathbb R}
\DeclareMathOperator{\Tr}{trace}
\DeclareMathOperator{\col}{col}
\DeclareMathOperator{\vect}{vec}
\DeclareMathOperator{\diag}{diag}
\newcommand{\lN}{\mleft\lVert}
\newcommand{\rN}{\mright\rVert}

\usepackage[varqu,scaled=0.95]{zi4}
\usepackage{listings}

\lstdefinelanguage{Rust}{
    morekeywords={
        as, async, await, break, const, continue, crate,
        dyn, else, enum, extern, false, fn, for, if,
        impl, in, let, loop, match, mod, move, mut,
        pub, ref, return, self, Self, static, struct,
        super, trait, true, type, unsafe, use, where,
        while
    },
    sensitive=true,
    morecomment=[l]{//},
    morecomment=[s]{/*}{*/},
    morestring=[b]",
}

\lstdefinelanguage{Verus}{
    language=Rust,
    morekeywords={
        requires, ensures, invariant, decreases,
        proof, spec, exec, ghost, tracked,
        open, closed, uninterp, broadcast, axiom,
        assume_specification,
        assert, assume, forall, exists,
        old, reveal, hide, by, recommends,
        opens_invariants, no_unwind
    },
    moredelim=[s][\color{gray}]{\#[}{]},
}

\lstdefinestyle{block}{
    basicstyle=\ttfamily\footnotesize,
    aboveskip=0.6\baselineskip,
    belowskip=0.6\baselineskip,
}

\begin{document}

\title{Formally Verified Semidefinite Programming}

\author{\IEEEauthorblockN{Perry Hart}
\IEEEauthorblockA{\textit{Institute for Computer-Aided Reasoning in Mathematics} \\
\textit{Carnegie Mellon University}\\
Pittsburgh, PA, USA \\
perryh@andrew.cmu.edu}
}

\maketitle

\begin{abstract}
We present a formally verified Rust implementation of a first-order semidefinite programming solver, a nontrivial piece of numerical software. We wrote the solver by hand and then directed Claude to generate the entire verification in Verus, an automated theorem prover for Rust programs. We discuss the algorithm underlying the solver, its implementation in Rust, and our pragmatic approach to proving it correct in Verus.
\end{abstract}

\begin{IEEEkeywords}
numerical optimization, semidefinite programming, Rust, formal verification, Verus, agentic AI
\end{IEEEkeywords}

\section{Introduction}
Semidefinite programming (SDP) is a powerful framework in convex optimization that sits atop a hierarchy of important conic programming classes, including second-order cone programming and, at the bottom, linear programming. Thanks to its expressive power, SDP has provided new computational techniques for combinatorial optimization~\cite{thetafunc,maxcup}, matrix completion~\cite{matcompl}, and quantum information theory~\cite{qitapp}, to name a few. Indeed, SDP is (numerically) solvable, up to a desired accuracy tolerance, in polynomial time by interior-point methods~\cite{ipmtheory}, which have given rise to open-source, performant SDP solvers~\cite{sedumi,sdpt3}. 

Interior-point methods, however, involve solving iteration-dependent Hessian systems---second-order information that scales with the number of linear constraints. As this number can be quadratic in the dimension of the decision variable, these methods are unsuitable for large-scale problems. To address this issue, researchers have developed first-order methods that reduce per-iteration cost (both time and memory) while maintaining global convergence. One prominent primal-dual method of this sort is due to Wen, Goldfarb, and Yin~\cite{WGY}, which we call the \emph{WGY method}. 

Of course, we would like assurance that the solution an SDP solver---a nontrivial piece of software---returns to us is correct. To this end, we present a self-contained, practical Rust implementation of the WGY method along with a proof of its correctness in Verus, an automated theorem prover for Rust programs~\cite{verusref}. To our knowledge, this is the first verified implementation of an SDP solver, thereby showing the community a feasible way to verify such software. We first wrote the Rust program manually, i.e., without agentic AI, and then systematically guided Claude (Fable 5.1) to produce the entire Verus proof. For this project, Claude proved to be adept at engineering robust and fast proofs.

The algorithmic core of our solver consists of a few fundamental numerical linear algebra routines. In particular, the workhorse of the WGY method is the spectral decomposition of a real symmetric matrix. For this purpose, our solver offers the user a choice between two variants of the (serial) Jacobi method: one with the traditional column-cyclic ordering of pivot entries and the other with a simple caching method to select the pivot of greatest magnitude.

To verify these routines with Verus, we must decide how to reason about floating-point computations, which permeate our solver. There are good reasons to avoid reasoning about floats directly. First, Verus currently offers little support for it. Second, the benefit of staying faithful to this aspect of the Rust code is outweighed by the additional complexity of floating-point error analysis. Indeed, this analysis would obfuscate what we're primarily after with Verus: a certified understanding of the code's \emph{mathematical} behavior, separate from artifacts of finite precision. Therefore, we decide to axiomatize a minimal set of basic floating-point operations as their exact real counterparts, for which Verus has extensive support. Within this framework, we verify the correctness of each linear algebra routine underpinning the solver.

On top of its algorithmic core, our solver provides an interface facilitating convenient formulations of SDP problems, which often involve splitting the decision variable into smaller ones. Our interface lets the user formulate the problem purely in terms of these composite variables. This decomposed problem is converted to an equivalent problem in standard form (with a single decision variable), which is fed to the WGY method.
Our proof of correctness relates the computed solution of the standard-form problem to the user-supplied problem, showing that the former is \emph{feasible}, i.e., satisfies the problem's constraints, in an approximate sense for the latter.  

Collectively, our Verus proofs  establish that the solver as a whole is correct: if it stops upon meeting its accuracy standard, it returns a primal-dual solution with small Karush--Kuhn--Tucker (KKT) residuals relative to the standard-form problem, in which case the primal solution is approximately feasible for the user-supplied problem. Since the objective functions of both problems are equal, these criteria make up a sensible notion of correctness for a numerical solver. Indeed, it is a basic theorem of convex optimization that a primal-dual solution exactly satisfying the KKT conditions (meaning that the residuals vanish) is optimal~\cite[Sec.~5.5.3]{BoydVand}.  

We do not try to formally verify convergence of our solver. Wen, Goldfarb, and Yin establish global convergence via convex analysis~\cite[Sec.~2.2]{WGY}, a theory for which Verus is ill-suited. Sacrificing a computer-checked proof of convergence for the benefits of Verus reflects the question \emph{we} sought to answer: when our Rust program stops with a claimed solution, should the user be confident in it?

\subsection*{Outline}

We consider our SDP solver and its Verus verification in turn. We begin with a brief overview of semidefinite programming (\cref{SDPbackg}). We then discuss the WGY method for solving semidefinite programs together with its implementation in Rust (\cref{RustWGY}). Afterward, we provide a brief overview of Verus (\cref{Verusbackg}) before describing what we verified about our solver (\cref{VerusSDP}). 
The codebase on which this paper is based, including the infrastructure built to steer Claude, is available at \cite{gitcode}.

\subsection*{Related Work}

The computer verification of numerical programs is an active research area. Its methods come in two main flavors: (a) formal proofs of correctness in a theorem prover, whether automated or interactive, and (b) certificate validation performed by a separate, trusted program. Flavor (a) has the advantage of specifying properties in an expressive formal system. It, however, involves defining the numerical program in question within this system, which may be far from the user's preferred programming language and may lead to worse performance. For example, proving correctness in Lean would require writing the executable code as a functional program, which will typically perform worse than an imperative program written in languages like C or Rust. Flavor (b) avoids this issue but shifts the burden of verification to the separate validator. Moreover, flavor (b) just tells us how to check a given output, not whether the code behaves as intended.

Both flavors of verification have been applied to convex optimization. For example, Li~et~al.\ have developed an extensive Lean library that formalizes the convergence rates of various gradient descent algorithms~\cite{jamrformfour}. Moreover, focusing on the solver interface, Bentkamp, Mir, and Avigad have built in Lean a verified translation from a user-supplied convex problem to one in  conic form (which is then sent to an external solver)~\cite{verreduction}. Both of these works, like ours, neglect floating-point reasoning by assuming exact real arithmetic---although Lean now offers more support than Verus for such reasoning~\cite{FloatLib}. Our work, however, targets a class of optimization problems not covered by gradient descent and addresses the end-to-end solver, not just the conversion step.

Meanwhile, certificate validation has had limited success in verifying SDP solvers. Primal-dual methods produce approximate solutions that can be directly checked against the KKT conditions. A validator will check each of these conditions with a trusted or, better, verified program, which ideally accounts for floating-point error. For a restricted class of SDP problems, Martin-Dorel and Roux devise a promising method for checking that the primal solution is positive semidefinite~\cite[Sec.~2.4.2]{refltacticpoly}. They implement an efficient decision procedure in Rocq and verify its soundness. Their method, though, is not applicable to a general solver like ours.  

Our project's approach fits into flavor (a): verification amounts to specifying and proving properties in Verus's formal system. It tries to prove that the computation of KKT residuals is correct, whereas a certificate validator merely tries to check that each particular result has small KKT residuals.

Ours appears to be the first verified implementation of an SDP solver. Surprisingly, it also appears to offer the first verified implementation of spectral decomposition for real symmetric matrices, although Lean's mathlib includes the spectral theorem, in noncomputable form, for Hermitian matrices~\cite[\href{https://leanprover-community.github.io/mathlib4_docs/Mathlib/LinearAlgebra/Matrix/Spectrum}{Spectral theory of Hermitian matrices}]{spthryherm}. Thus, the present paper marks an important first step in research on verified numerical optimization.

\section{Background on SDP} \label{SDPbackg}

Semidefinite programming seeks to minimize a real-valued linear objective over the convex cone of positive semidefinite (real) matrices subject to finitely many linear constraints:
{
\renewcommand{\theequation}{\textsf{std}}
\begin{mini}|l|
  {X \in \S_n^+}{C \bullet_F X}{\label{stdform}}{}
  \addConstraint{}{A_i \bullet X= b_i \quad}{i = 1, \ldots, m}
 \end{mini}%
}%
This presentation is the \emph{standard form} of a semidefinite program (SDP). Here, $\S_n^+$ denotes the set of positive semidefinite $n$-by-$n$ matrices: a square matrix $M$ is \emph{positive semidefinite (PSD)} if it is symmetric, i.e., an element of the $\frac{n\mleft(n+1\mright)}{2}$-dimensional vector space $\S_n$, and $x^TMx \geq 0$ for all $x \in \R^n$. The latter condition is equivalent to requiring $M$'s eigenvalues, which belong to $\R$, to be nonnegative. The parameter matrices $C$ and $A_i$ are assumed, without loss of generality, to be symmetric. The operation $A_i \bullet X$ is defined as the ordinary dot product of $A_i$ and $X$ viewed as vectors of size $\frac{n\mleft(n+1\mright)}{2}$. A variant of this operation is the \emph{Frobenius inner product} $C \bullet_F X$, defined by considering $C$ and $X$ as $n \times n$ matrices and taking their dot product in $\R^{n^2}$. Another way of writing the linear constraints is $\A(X) = b$ where $\A$ has rows $A_1, \ldots, A_m$.

A more flexible presentation lets one decompose the SDP into multiple decision variables. To describe it, we need the notion of \emph{linear matrix inequality (LMI)}: a relation $M_1 \succeq M_2$ on matrices defined to hold when $M_1 - M_2$ is positive semidefinite---in particular, $M \succeq 0$ means $M \in \S_n^+$. Now, the \emph{convenient} presentation of an SDP has the form
{
\renewcommand{\theequation}{\textsf{split}}
\begin{mini}|l|
  {X_1, \ldots, X_k}{C_1 \bullet_F X_1 + \cdots + C_k \bullet_F X_k}{\label{splitform}}{}
  \addConstraint{}{X_i \in \S_{d_i} \quad}
    {i = 1, \ldots, k}
  \addConstraint{}{\textstyle\sum_{i=1}^k A_{j,i} \bullet X_i = v_j \quad}
    {j = 1, \ldots, m}
  \addConstraint{}{\textstyle\sum_{i=1}^k \mathcal{L}_{t,i}(X_i) \succeq M_t \quad}
    {t = 1, \ldots, \ell}
\end{mini}%
}%
where each $\L_{t,i}$ is a linear map $\S_{d_i} \to \mathsf{Mat}_{n_t}(\R)$ into the space of $n_t \times n_t$ matrices for some $n_t \geq 0$, with $d_i$ the dimension of the decision variable $X_i$. Often, it is far easier to formulate an SDP with multiple decision variables than with just one. In fact, we believe that to be useful software, an SDP solver \emph{must} accommodate the convenient presentation.
\begin{example}
The \emph{nuclear norm} of a matrix is the sum of its singular values. Minimizing it over a constraint set is an important problem in data science and can be formulated as the following SDP~\cite{siamRechtFazelParrilo}:
{
\begin{mini*}|l|
  {X, W_1, W_2}{\frac{1}{2}\mleft(\Tr(W_1) + \Tr(W_2)\mright)}{}{}
  \addConstraint{}{A(X) = b}
  \addConstraint{\begin{pmatrix} 
  W_1 & X \\
  X^T & W_2 
  \end{pmatrix} \succeq 0}
 \end{mini*}%
}%
\end{example}

Every convenient presentation can be transformed into an equivalent standard-form problem. We call this process \emph{canonization}, which rests on the following standard fact.

\begin{lemma} \label{blockdec}
Let $B_1, \ldots , B_k$ be square matrices (of possibly different sizes). The block-diagonal matrix 
\[
\begin{pmatrix}
B_1 & & \\ & \ddots & \\ & & B_k 
\end{pmatrix}
\] is PSD if and only if each block is.
\end{lemma}

With \cref{blockdec}, we can assemble the decision variables $X_1, \ldots, X_k$ into a big block-diagonal one as follows. Every symmetric matrix $S$ is the difference $P_1 - P_2$ of two PSD matrices: take $\mleft(S + rI\mright) - rI$ with $r \coloneqq \lN{S}\rN_{\infty}$, which is at least the spectral radius of $S$ because $\lN{\cdot}\rN_{\infty}$ is sub-multiplicative. Therefore, the set of feasible solutions for \eqref{splitform} corresponds to the set of $X \succeq 0$ such that
\begin{align*}
& X = 
\begin{pmatrix}
U_1 \\
& \ddots \\
& & U_{\ell} \\ 
& & & P_{1,1} \\
 & & &  & P_{1,2} \\
 & & & & & \ddots \\
 & & & & & & P_{k,1} \\
 & & & & & & & P_{k,2}
\end{pmatrix} \\
& 
\mathcal{L}_{t,1}(P_{1,1} - P_{1,2}) + \cdots + \mathcal{L}_{t,k}(P_{k,1} - P_{k,2}) - U_t = M_t \\
& A_{j,1} \bullet \mleft(P_{1,1} - P_{1,2}\mright)  + \cdots + A_{j,k} \bullet \mleft(P_{k,1} - P_{k,2}\mright) = v_j 
\end{align*}
We call this $X$ the \emph{lift} of $X_1, \ldots, X_k$. For each $i \in \mleft\{ 1, \ldots, k\mright\}$, we recover the decision variable $X_i$ as $P_{i,1} - P_{i,2}$. The constraints imposed on $X$ to give it the desired shape are all linear, and the expression
\[
C_1 \bullet_F \mleft(P_{1,1} - P_{1,2}\mright) + \cdots + C_k \bullet_F \mleft(P_{k,1} - P_{k,2}\mright)
\] defines a real-valued linear function of $X$. It follows that minimizing this map over all $X \succeq 0$ satisfying these constraints not only solves \eqref{splitform} but also can readily be put in standard form. We define the canonization of \eqref{splitform} as the resulting standard-form problem.

We now know that it suffices to solve \eqref{stdform}, but how do we know when we have solved it? To answer this, we view \eqref{stdform} as the \emph{primal} problem and consider its \emph{dual}:
{
\renewcommand{\theequation}{\textsf{dual}}
\begin{maxi}|l|
  {y, Z}{b \bullet y}{\label{dualform}}{}
  \addConstraint{}{\A^{\ast}(y) + Z = C}
  \addConstraint{}{Z \in \S_n^+}
 \end{maxi}%
}%
Here, the map $\A^{\ast} : \mleft(\R^m, \bullet\mright) \to \mleft(\S_n, \bullet_F\mright)$ denotes the \emph{adjoint} of $\A$ viewed as a map of inner product spaces, characterized by the property that $\A(X) \bullet y = X \bullet_F \A^{\ast}(y)$ for all $X \in \S_n$ and $y \in \R^m$. Now, consider the \emph{KKT conditions} on $\mleft(X,y,Z\mright)$:
\begin{itemize}
\item $X$ is feasible for \eqref{stdform}.
\item $\mleft(y,Z\mright)$ is feasible for \eqref{dualform}.
\item $C \bullet_F X  = b \bullet y$.
\end{itemize} 
These conditions are simply the general KKT conditions (which apply to virtually any optimization problem) specialized to the SDP setting~\cite[Sec.~3.1]{VandBoydsiamreview96}. 
As the objective and constraints of \eqref{stdform} are convex, if $\mleft(X,y,Z\mright)$ satisfies the KKT conditions, then $X$ and $\mleft(y,Z\mright)$ are optimal solutions for the primal and dual, respectively. Conversely, the KKT conditions are necessary for a triple to be optimal provided that both the primal and the dual satisfy \emph{Slater's condition}~\cite[Sec.~5.2.3]{BoydVand}: there exist feasible solutions $X, y, Z$ such that both $X$ and $Z$ are \emph{positive definite}, i.e., have positive eigenvalues.

A \emph{primal-dual} solver is one that computes solutions to the primal and dual simultaneously. Convergence of a plain primal-dual solver, such as the WGY method, to a triple $\mleft(X,y,Z\mright)$ satisfying the KKT conditions relies on Slater's condition, which generally must be checked by hand.

Since, in practice, we must stop an SDP solver after finitely many iterations, we measure the accuracy of its returned solution $\mleft(X,y,Z\mright)$ by the \emph{KKT residuals}: quantities measuring how close $\mleft(X,y,Z\mright)$ is to satisfying the KKT conditions. Small residuals correspond to high accuracy. In practice,  the user may forget Slater's condition and simply test whether the solver returns small KKT residuals. It is this realistic scenario that will shape what we want to prove about our solver.

\section{Rust Implementation of WGY Method} \label{RustWGY}

In this section, we describe how we implement in Rust a complete SDP solver based on the WGY method. We do not replay the entire algorithm, which is explained in \cite[Sec.~2.1]{WGY}. Rather, we sketch it while focusing on those steps which, as we'll see, play primary roles in the verification.

\subsection*{Canonization}

The user supplies their SDP in convenient form. Its canonization is implemented like so:
\begin{lstlisting}[style=block,
    language=Rust
]
fn sdp_to_standard(sdp : &SDP, check_cns : bool) ->
  (SymMatrix, (Matrix, Vec<(usize, usize)>), Vector, ...)
\end{lstlisting}
Here, \lstinline[language=Rust]|SDP| is a struct whose fields specify the parameters of a convenient presentation. For this discussion, the following are the relevant components of the output tuple, from left to right.
\begin{itemize}
\item The symmetric matrix $C$ defining the objective map. The type of $C$, \lstinline[language=Rust]|SymMatrix|, is an alias for \lstinline[language=Rust]|Vector|, which is in turn an alias for \lstinline[language=Rust]|Vec<f64>|. A vector stores the upper triangle of a symmetric matrix by flattening its columns.
\item The constraint matrix $\A$ represented as a pair $\mleft(M, \zeta\mright)$. The type of $M$, \lstinline[language=Rust]|Matrix|, is another alias for \lstinline[language=Rust]|Vector|, again storing a matrix by flattening its columns. The function computes $M$ as the matrix $\Sigma$ of the linear map
\begin{multline*}
X \mapsto \col\bigl(\vect L_1(X), \ldots, \vect L_\ell(X), \\
N_1(X), \ldots, N_m(X)\bigr)
\end{multline*}
where $\col(\cdot)$ denotes vertical concatenation, $\vect(\cdot)$ flattens a matrix column-wise, and $L_t(X) \coloneqq \mathcal{L}_{t,1}(P_{1,1} - P_{1,2}) + \cdots + \mathcal{L}_{t,k}(P_{k,1} - P_{k,2}) - U_t$ and $N_j(X) \coloneqq A_{j,1} \bullet \mleft(P_{1,1} - P_{1,2}\mright)  + \cdots + A_{j,k} \bullet \mleft(P_{k,1} - P_{k,2}\mright)$, as in \cref{SDPbackg}. For efficiency, the zero constraints making $X$ block-diagonal are stored separately in $\zeta$, which consists of all row-column indices where $X$ must be zero.
\item The right-hand side $b_{\Sigma}$. The function computes it as $\col\mleft(\vect M_1, \ldots, \vect M_\ell, v\mright)$. For efficiency, the right-hand sides of the zero constraints are left implicit, but the full right-hand side for $\A$ is $b \coloneqq \col\mleft(b_{\Sigma}, 0_{\lvert \zeta \rvert}\mright)$.
\end{itemize}

The function's second argument, \lstinline[language=Rust]|check_cns|, is a flag that tells the solver to check whether the augmented matrix $\mleft[\A \mid b\mright]$ has redundant constraints (after checking that it's consistent in the first place) and, if so, to remove them. This one-time check determines whether $\A$ has full rank and, if not, trims it so that it does. The WGY method requires $\A$ to have full rank, so that its Gram matrix $\A\A^{\ast}$ is invertible. If the user knows beforehand that the constraints are non-redundant, they may forgo the check.   

\subsection*{Rank Computation}

If \lstinline[language=Rust]|check_cns| is turned on, then we need to compute the rank of $\A$. To do so, we use Gaussian elimination (with partial pivoting), implemented like so:
\begin{lstlisting}[style=block,
    language=Rust
]
fn gauss_elim(mat: &mut Matrix, col_dim: usize, row_dim: usize) -> (&mut Matrix, usize, bool)
\end{lstlisting}
This function mutates the input matrix to place it in row echelon form. Along the way, it tracks the number of valid pivots in $\A$ and thereby computes the matrix's rank, returning it as the second component. It also returns a flag that indicates whether the matrix, when its final column is viewed as the right-hand side of a linear system, is consistent.

The Gaussian elimination routine illustrates the kind of numerical reasoning that we will be forced to handle. In particular, we skip pivots using the following criterion:
\begin{lstlisting}[style=block,
    language=Rust
]
if pivot_abs == 0.0 || pivot_abs < TOL * coeff_scale { continue; }
\end{lstlisting}
Here, \lstinline[language=Rust]|TOL| is a small positive constant, and \lstinline[language=Rust]|coeff_scale| is the maximum magnitude over all entries in $\A$, so that their product is a scale-invariant threshold.
With ordinary real numbers, a correct Gaussian elimination routine skips precisely those pivots equaling zero on the nose. The reality of floating-point arithmetic forces us to use a tolerance-based test instead. Thus, we cannot hope to prove exact versions of correctness properties in Verus, even when the basic framework assumes that floating-point operations are exact.

\subsection*{Standard-Form Solver}

Once the preprocessing is done, we run the WGY method on the data $\mleft(C, \A, b\mright)$. This method rests on an auxiliary function called the \emph{augmented Lagrangian}:
\begin{align*}
& \Phi_{\mu}\mleft(X,y,Z\mright) \  \coloneqq \  \mathop{-b \bullet y} + X \bullet_F \Delta + \frac{\lN{\Delta}\rN_F^2}{2\mu}
\\ & \Delta \ \coloneqq \  \A^{\ast}(y) + Z - C
\end{align*} for a given penalty parameter $\mu > 0$.
At each step, the WGY method minimizes $\Phi_{\mu}$ over $y \in \R^m$ and then over $Z \in \S_n^+$, thereby serving as an \emph{alternating direction} method (which is why the name \emph{sdpad} appears in our codebase). It then updates $X$ in terms of the computed minimizers. 

In this discussion, we focus on the iterative computations of $y$ and $Z$. The update formula for the former follows from the stationarity condition $\nabla_y{\Phi_{\mu}} = 0$ (a general KKT condition):
\[
\A\A^{\ast}y^{k+1} = \mu\mleft(b - \A\mleft(X^k\mright)\mright) + \A\mleft(C - Z^k\mright) 
\] where $\A\A^{\ast}$ is positive definite, hence invertible, because $\A$ has full rank. We thus need an algorithm for solving positive definite linear systems. 

To this end, we implement $LDL^T$ decomposition of an $n \times n$ symmetric matrix $S$, an efficient factorization that succeeds if and only if $S$ is positive definite (assuming exact arithmetic):
\begin{lstlisting}[style=block,
    language=Rust
]
fn ldlt_decomp(mat : &SymMatrix, dim : usize) -> Option<(LowTriMatrix, Vector)> 
\end{lstlisting}
(Here, \lstinline[language=Rust]|LowTriMatrix| is another alias for \lstinline[language=Rust]|Vector|.)
By decomposing $S$, a system $Sx = v$ becomes a composite of triangular solves. With the exact-real framework used by our verification, we can prove that when \lstinline[language=Rust]|ldlt_decomp(S, n)| returns a pair $\mleft(L, D\mright)$, we have $LDL^T = S$ on the nose.  Nevertheless, when we use this routine to test whether a matrix is positive definite (as we'll do in our stopping criteria), we pick up numerical error. Indeed, \lstinline[language=Rust]|ldlt_decomp| tests positivity of the diagonal entries with
\begin{lstlisting}[style=block,
    language=Rust
]
if d[i] < TOL * max_diag_sym(mat, dim).max(1.0) { return None; }
\end{lstlisting}

\begin{remark}
This stage of the WGY method shows why we keep the zero constraints separate from those encoded by $\Sigma$. The latter only involve entries within $X$'s diagonal blocks, so $\A\A^{\ast}$ is block-diagonal with blocks $\Sigma\Sigma^{\ast}$ and $\frac{1}{2}I$ (where the factor $\frac{1}{2}$ reflects how $\bullet_F$ differs from $\bullet$). Hence we need only run $LDL^T$ on $\Sigma\Sigma^{\ast}$, which is typically far smaller than $\A\A^{\ast}$.
\end{remark}

The update formula for $Z$ comes from noticing that minimizing $\Phi_{\mu}(X^k, y^{k+1}, Z)$ over $Z$ is equivalent to the problem
\begin{mini*}|l|
  {Z \in \S_n^+}{\lN{Z - V^{k+1}}\rN_F^2}{}{}
 \end{mini*}%
where $V^{k+1}$ denotes the symmetric matrix $C - \A^{\ast}(y^{k+1}) - \mu X^k$. This problem has a known closed-form solution~\cite[Sec.~2]{higham88}: the nonnegative part of $V^{k+1}$'s spectral decomposition. Therefore, we want a spectral decomposition algorithm. 

To this end, we implement two variants of the \emph{Jacobi method}~\cite[Sec.~8.5]{golubvanloan}.
Let $S \in \S_n$. The Jacobi method works by iteratively choosing a nonzero off-diagonal entry $\mleft(p,q\mright)$ of $S$, called the \emph{pivot}, and making it zero by applying an $O\mleft(n\mright)$-time orthogonal transformation $J_{p,q}$, called a \emph{Jacobi rotation}. The method stops when all off-diagonal entries are small enough. Under this method, $S$ converges quadratically to the diagonal of its eigenvalues~\cite[Sec.~5.3.5]{appliednumla97}. 

One has a choice of how to choose the pivot on each iteration. The standard option is to employ column-cyclic ordering, which simply traverses the columns and takes the current entry as the next pivot. This approach has the advantage of making pivot selection constant time. It does, however, require calling a Jacobi rotation on every off-diagonal entry. 

Another option is to dynamically select the largest entry (in magnitude) as the next pivot. Doing a brute-force search over $S$ for this entry is too expensive. Instead, the program maintains a cache $\pi$ that stores, for each column index $c \geq 1$, the row index $\pi_c < c$ of the largest entry above the diagonal in column $c$. The transformation $J_{p,q}$ only modifies entries landing either in the $p$-th or $q$-th row or in the $p$-th or $q$-th column. So, to update $\pi_c$, one just needs to check how the rotation affected rows $p$ and $q$ in column $c$:
\begin{enumerate}
\item If $c \in \mleft\{p,q\mright\}$, then scan the $c$-th column to find its new largest entry.
\item Otherwise, check whether the current $\pi_c$ equals $p$ or $q$. If no, then just compare $S_{\pi_c,c}$ with $S_{p,c}$ and $S_{q,c}$ and choose the row index of the largest entry.
\item In the event that $\pi_c \in \mleft\{p, q\mright\}$, which occurs with probability $O\mleft(\frac{1}{n}\mright)$, check whether $\lvert{S_{\pi_c,c}}\rvert$ fell below the magnitude of the old pivot. If no, then $S_{\pi_c,c}$ is still at least as big as each untouched entry of column $c$, so just compare $S_{p,c}$ and $S_{q,c}$ and choose the row index of the larger entry. If yes, then scan the $c$-th column to find its new largest entry.
\end{enumerate}  
With this scheme, pivot selection still takes quadratic worst-case time but takes only linear time on average. This approach has the advantage of taking possibly fewer \emph{sweeps} over $S$---i.e., passes of $\frac{n\mleft(n-1\mright)}{2}$ rotations---than the column-cyclic version, but pivot selection no longer takes constant time.

We have implemented both approaches:
\begin{lstlisting}[style=block,
    language=Rust
]
fn jacobi_eigen_cyc(a : &mut SymMatrix, dim : usize)
    -> (Vector, Matrix)
fn jacobi_eigen_cach(a : &mut SymMatrix, dim : usize)
    -> (Vector, Matrix)
\end{lstlisting}
These routines store the diagonal of eigenvalues in the \lstinline[language=Rust]|Vector| output and the eigenvectors in the \lstinline[language=Rust]|Matrix| output.
The user decides which one the SDP solver calls.

\subsection*{Stopping Criteria}

While the WGY method converges on paper (assuming Slater's condition), we need criteria for deciding when to actually stop the solver. We follow the criteria  from \cite[Sec.~3.4]{WGY}. Essentially, the method is deemed to have converged when the KKT residuals for an iterate $\mleft(X^k, y^k, Z^k\mright)$ are sufficiently small. Here, the hardest condition to check is that $X^k$ is nearly PSD. We use a popular heuristic for this test: choose a small value $\varepsilon > 0$ and run $LDL^T$ on $X^k + \varepsilon I$. This approach introduces additional numerical error to our solver but is relatively fast.

\section{Background on Verus} \label{Verusbackg}

Why did we choose Verus? We chose Rust because it compiles programs to fast machine code, and we don't want to sacrifice performance to formally check that our solver's implementation is correct. Thus, Verus is the right tool for us because it enables the \emph{direct} static verification of our Rust solver, without having to transpile it into a different language. Verus takes a large subset of Rust and treats it as a proof language for its own executable programs. It does so by extending the Rust compiler so that it associates exactly one of three modes to every Rust expression: \emph{executable}, \emph{specification}, or \emph{proof}. Verus compiles only executable code to machine code, thereby keeping the runtime performance of our solver from deteriorating.
 
Verus augments Rust's type system with types like $\mathbb{Z}$ and $\mathbb{R}$ for specification and proof modes. In particular, specifications are formulas in the language of an SMT theory and are used as pre- and postconditions (marked by \lstinline[language=Verus]|requires| and \lstinline[language=Verus]|ensures|, respectively) in functions written in proof or executable mode. Within such functions, the user supplies, where appropriate, loop invariants as specifications. At compile time, Verus translates specification code to SMT queries and calls Z3~\cite{demourabjorner08} on them. To aid Z3, the user supplies separate lemmas (which are assigned proof mode) as hints.

In general, the goal is to prove that an executable Rust function satisfies its postcondition under its precondition. As is the case with Hoare-style reasoning, though, such proofs are \emph{partial}: each postcondition holds when the function returns normally, but whether it necessarily does so is outside the assertion's scope. For example, we do not analyze whether a \lstinline[language=Rust]|panic!| in our solver is reachable. 

\section{Verification of Rust SDP Solver} \label{VerusSDP}

In this section, we report on how we verified the solver described in \cref{RustWGY}. We steered Claude to generate the entire verification in Verus, finding it highly adept at engineering proofs of individual specifications. The one aspect on which we needed to provide oversight was the design of the specifications themselves (although Claude was usually proficient at it). Initial ambiguities in our prompts gave Claude too much freedom in designing them, causing some of them to be weaker than intended. We reviewed Claude's work in stages (the nodes of a dependency graph we prompted Claude to create), sometimes with the help of a different agent, namely Codex. As we were familiar with the executable code and with what we wanted to prove about it, we had little trouble steering Claude to fix its problematic specifications. 

In total, Claude wrote almost 18,000 logical lines of Verus. It is impractical to record here every important theorem in them. Rather, we sketch how this collection of guarantees comes together by walking the reader through a handful of its essential features.

\subsection*{Trusted Base}

The trusted base establishes our approach to floating-point reasoning, namely to sidestep it. Formally, it axiomatizes floating-point arithmetic as exact real arithmetic. Importantly, this approach lets one reason about our solver's floating-point computations via Z3's powerful nonlinear arithmetic solver. 

First, we introduce the uninterpreted function \lstinline[language=Verus]|rv|, which gives the exact real value of an \lstinline[language=Rust]|f64|. We then axiomatize the behaviors of basic operations like addition, comparison, and Rust's native absolute value method:
\begin{lstlisting}[style=block,language=Verus]
uninterp spec fn rv(x: f64) -> real;

broadcast axiom fn axiom_add(a: f64, b: f64, o: f64)
    requires #[trigger] add_ensures::<f64>(a, b, o),
    ensures rv(o) == rv(a) + rv(b);

broadcast axiom fn axiom_lt(a: f64, b: f64, o: bool)
    requires #[trigger] lt_ensures::<f64>(a, b, o),
    ensures o == (rv(a) < rv(b));

assume_specification[ f64::abs ](x: f64) -> (r: f64)
    ensures rv(r) == abs_r(rv(x));
\end{lstlisting}
Here, \lstinline[language=Verus]|add_ensures| and \lstinline[language=Verus]|lt_ensures| are uninterpreted predicates from Verus's standard library.

\subsection*{Matrices}

Of course, to reason about our solver, we need to handle matrix computations. Claude chose to represent a matrix as a function from index pairs to reals and to read matrices stored as packed vectors---represented as sequences of reals in Verus---as such functions:
\begin{lstlisting}[style=block,language=Verus]
type MatF = spec_fn(int, int) -> real;

spec fn sym_view(m: Seq<real>) -> MatF {
    |r: int, c: int| if r <= c { m[pk_idx(r, c)] }
                     else { m[pk_idx(c, r)] } }
\end{lstlisting}
Here, \lstinline[language=Verus]|pk_idx(r, c)| is the position of entry $\mleft(r,c\mright)$ in the packed upper triangle \lstinline[language=Verus]|m|. Thus, \lstinline[language=Verus]|sym_view(m)| is symmetric by construction. To handle matrices from LMI constraints, the function \lstinline[language=Verus]|cm_view(m, dim)|, not shown, similarly reads a general column-major matrix with \lstinline[language=Verus]|dim| rows.

\subsection*{Spectral Decomposition}

A correct spectral decomposition algorithm, or \emph{eigensolver}, that returns $\mleft(Q,B\mright)$ on symmetric input $A$ ensures that $A = Q B Q^T$ with $B$ diagonal and $Q$ orthogonal. For our implementation of the Jacobi method, we can prove, thanks to our trusted model, that $A = Q B Q^T$ on the nose. We can also prove that $Q$ is orthogonal. But since we do not test zero off-diagonal entries exactly, we merely ask for $B$ to be close to its own diagonal. Such a residual is a reasonable quantity to control. Indeed, since the eigensolver keeps $A$ orthogonally similar to $B$ and $B = \diag(B) + \mleft(B - \diag(B)\mright)$, the Hoffman--Wielandt inequality~\cite[Thm.~VI.4.1]{BhatiaMatrixAn97} implies that
\[
\lN{\lambda(A) - \lambda(\diag(B))}\rN \ \leq \ \lN{B - \diag(B)}\rN_F
\] where $\lambda(\cdot)$ denotes the vector of eigenvalues in decreasing order. By bounding this Frobenius norm, we bound how much the eigenvalues returned by the solver, namely the diagonal of $B$, collectively deviate from $A$'s true eigenvalues.

Based on the preceding discussion, we formulate in Verus what it means for an eigensolver's output to be correct:
\begin{lstlisting}[style=block,language=Verus]
spec fn eigen_decomp_inv(
    a: Seq<real>, q: Seq<real>, b: Seq<real>, dim: int)
    -> bool {
    &&& is_orthogonal(cm_view(q, dim), dim)
    &&& meq(sym_view(a),
            mmul(mmul(cm_view(q, dim), sym_view(b), dim),
                 mtr(cm_view(q, dim)), dim),
            dim, dim) }

spec fn eigen_decomp_approx(
    a: Seq<real>, q: Seq<real>, b: Seq<real>,
    dim: int, eps: real) -> bool {
    &&& eigen_decomp_inv(a, q, b, dim)
    &&& offdiag_norm_r(b, dim) <= eps }

spec fn eigensolver_ok(
    tol: real, a: Seq<real>, b: Seq<real>, q: Seq<real>,
    vals: Seq<real>, dim: int) -> bool {
    &&& q.len() == dim * dim
    &&& vals == sym_diag_r(b, dim)
    &&& eigen_decomp_approx(a, q, b, dim,
            jacobi_eps_r(tol, b, dim)) }
\end{lstlisting}
Here, \lstinline[language=Verus]|jacobi_eps_r(tol, b, dim)| is the value
$\sqrt{n\mleft(n-1\mright)} \cdot \mathsf{tol} \cdot \max\mleft(\max_i \lvert{B_{ii}}\rvert, 1\mright)$
with $n$ the matrix dimension. This bound is derived from the explicit stopping test in the executable code: the eigensolver stops once all off-diagonal entries of $B$ have magnitude below $\mathsf{TOL} \cdot \max\mleft(1, \max_i \lvert{B_{ii}}\rvert\mright)$.

We prove that both variants of the Jacobi method---the \lstinline[language=Verus]|exec| functions \lstinline[language=Verus]|jacobi_eigen_cyc| and \lstinline[language=Verus]|jacobi_eigen_cach|---are correct eigensolvers in this sense. Moreover, we verify the distinguishing feature of each one. For the column-cyclic variant, we prove that a sweep visits every off-diagonal entry exactly once:
\begin{lstlisting}[style=block,language=Verus]
spec fn visits_each_pair_once(
    visited: Seq<(int, int)>, dim: int) -> bool {
    &&& forall|j: int, k: int| 0 <= j < k < dim ==>
            0 <= #[trigger] cyc_pos(j, k) < visited.len()
            && visited[cyc_pos(j, k)] == (j, k)
    &&& forall|t: int| 0 <= t < visited.len() ==>
            0 <= (#[trigger] visited[t]).0 < visited[t].1 < dim
            && t == cyc_pos(visited[t].0, visited[t].1) }
\end{lstlisting}
For the caching variant, we prove that the cache $\pi$, as in \cref{RustWGY}, stores what it's supposed to throughout \lstinline[language=Verus]|jacobi_eigen_cach|'s main loop. Formally, \lstinline[language=Verus]|col_max_at(b, c, i)| expresses that row $i < c$ holds an entry of maximal magnitude among the entries above the diagonal in column $c$, and we prove that $\pi$ satisfies the following predicate:
\begin{lstlisting}[style=block,language=Verus]
spec fn cache_ok(
    b: Seq<real>, maxima: Seq<usize>, dim: int) -> bool {
    &&& maxima.len() == dim - 1
    &&& forall|t: int| 0 <= t < dim - 1 ==>
            col_max_at(b, t + 1,
                       #[trigger] maxima[t] as int) }
\end{lstlisting}

\subsection*{KKT Residuals}

Suppose that the solver returns $\mleft(X, y , Z\mright)$ claiming that it is a primal-dual solution to the \emph{canonical problem}, the canonization of the user-supplied problem. Our verification's central result is that the convenient-form points $X_1, \ldots, X_k$ extracted from $X$ make up an approximate solution to the user-supplied problem. To achieve it, we verify the following two properties:
\begin{enumerate}
\item The returned triple has small KKT residuals relative to the canonical problem.
\item Any approximately feasible point for the canonical primal problem induces approximately feasible points for the user-supplied problem.
\end{enumerate}

To check property (1), we in particular want to verify that the primal solution claimed by the solver is (approximately) PSD. The primal variable keeps its block-diagonal form at each iteration, so to check that it is PSD, we check that each block is. Recall that to do so, we use the heuristic that checks the predicate \lstinline[language=Verus]|psd_shifted(a, dim, eps)| defined as $x^T A x + \varepsilon \lN x \rN^2 \geq 0$ for all $x$. The predicate \lstinline[language=Verus]|block_psd_ok| applies this test to each diagonal block $B$---which is symmetric by construction and specified by a starting position and a dimension---with $\varepsilon$ set to $\mathsf{TOL} \cdot \max\mleft(\max_i \lvert{B_{ii}}\rvert, 1\mright)$:
\begin{lstlisting}[style=block,language=Verus]
spec fn psd_shifted(a: MatF, dim: int, eps: real)
    -> bool {
    forall|x: Seq<real>| x.len() == dim ==>
        #[trigger] quad_r(a, dim, x) + eps * dot_r(x, x)
            >= 0.0real }

spec fn block_psd_ok(
    tol: real, a: Seq<real>, start: int, dim: int) -> bool {
    psd_shifted(block_view(a, start), dim,
                psd_check_eps_r(tol, a, start, dim)) }

fn psd_block_check(
    mat: &SymMatrix, blocks: &Vec<(usize, usize)>)
    -> (result: bool)
    requires blocks_fit(blocks@, mat@.len()),
    ensures
        result ==> forall|b: int| 0 <= b < blocks@.len() ==>
            #[trigger] block_psd_ok(rv(TOL), rvs(mat@),
                blocks@[b].0 as int, blocks@[b].1 as int),
\end{lstlisting}
The rest of property (1) is easier: the dual variable always remains PSD by construction, and all linear residuals are routine computations.

For property (2), we prove a more general relation between the user-supplied problem and the canonical one with the following soundness lemma. We just consider the case where the rank check is turned on as the other case is simpler:
\begin{lstlisting}[style=block,language=Verus]
proof fn lemma_approx_sound_reduced(
    sdp: &SDP, s: &SdpadSetup, z: Seq<real>)
    requires
        sdpad_input_ok(sdp, true),
        sdpad_setup_ok(sdp, true, *s),
        sdp.lmi@.len() >= 1,
        z.len() == pk_len(s.var_dim as int),
    ensures (...)
\end{lstlisting}
Here, \lstinline[language=Verus]|SdpadSetup| consists of standard-form data that the solver feeds to the WGY method and \lstinline[language=Verus]|sdpad_setup_ok| is the predicate that this data is the canonized user-supplied problem.

As desired, this soundness lemma ensures that canonization reflects feasibility: given any point $X$ for the canonical problem, the convenient-form points $X_1, \ldots, X_k$ violate the user-supplied constraints by an amount partly controlled by the \emph{primal infeasibility $\mathsf{pinf}$} at $X$, i.e., how much $\A(X)$ (after reduction) deviates from $b$. In more detail, let $\mathsf{linr}$ be the amount by which $X$ violates the \emph{unreduced} system $\mleft[A_0 \mid b_0\mright]$, before redundant rows were removed. Then, letting $\varphi(X)\coloneqq \lN A_0 \rN_F \lN X \rN + \lN b_0 \rN$, \lstinline[language=Verus]|lemma_approx_sound_reduced| ensures that
\[ 
\mathsf{linr} \ \leq \ \Gamma \cdot \mathsf{TOL} \cdot \varphi(X) + \lN E \rN_F \cdot \mathsf{pinf} \cdot \mleft(1 + \lN b \rN\mright)
\]
where the quantities $\Gamma$ and $E$ depend only on $\mleft[A_0 \mid b_0\mright]$ and the row echelon form that Gaussian elimination computes from it, not on $X$. Note that $\varphi(X)$ is the scaling factor in the classical normwise backward error for $\mleft[A_0 \mid b_0\mright]$~\cite{rigalgaches67}, so its appearance is expected. The bound  on $\mathsf{linr}$ bounds each linear residual $\lN{\sum_i A_{j,i} \bullet X_i - v_j} \rN$ as well as each LMI residual $\lN R_t \coloneqq \mleft(\sum_i \L_{t,i}(X_i) - M_t\mright) - U_t \rN_F$ (with notation as in \cref{SDPbackg}). Further, if the block $U_t$ passes the PSD check with shift $\varepsilon_t$, then soundness ensures that for all $x$,
\[
x^T \mleft(\textstyle\sum_i \L_{t,i}(X_i) - M_t\mright) x \ \geq \ \mathop{-\mleft(\varepsilon_t + \lN R_t \rN_F\mright) \lN x \rN^2}
\]
The smallest eigenvalue of a symmetric matrix $S$ is $\min_{x \neq 0} \frac{x^T S x}{ \lN x\rN^2}$, so this inequality implies that every eigenvalue of $\sum_i \L_{t,i}(X_i) - M_t$ (really, of its symmetric part) is at least $\mathop{-\mleft(\varepsilon_t + \lN R_t\rN_F\mright)}$, which shrinks as the solver drives down $\mathsf{pinf}$.

Finally, for our central result to be legitimate, canonization must not introduce error in the objective function. The soundness lemma ensures that it does not: the standard-form objective at $X$ equals the user-supplied one at $\mleft(X_1, \ldots, X_k\mright)$.

It remains to combine, in Verus, properties (1) and (2) to get our central result. Claude does so with the lemma \lstinline[language=Verus]|lemma_outcome_sound|, which applies the soundness lemma to the iterate the solver claims satisfies property (1). This claim is encoded in the predicate \lstinline[language=Verus]|outcome_ok|. Claude verifies that the solver's claim is true (roughly, that the KKT residuals are below $\mathsf{TOL}$), thereby deducing the following contract for the executable solver:
\begin{lstlisting}[style=block,language=Verus]
fn sdpad_verified(sdp: &SDP, check: bool,
    var: &JacobiVariant) -> (o: SdpadOutcome)
    requires sdpad_input_ok(sdp, check),
    ensures exists|s: SdpadSetup|
        sdpad_setup_ok(sdp, check, s)
        && outcome_ok(sdp, &s, check, o)
        && user_outcome_ok(sdp, check, &s, o),
\end{lstlisting}
Here, \lstinline[language=Verus]|user_outcome_ok|, proved with \lstinline[language=Verus]|lemma_outcome_sound|, states that the returned convenient-form points $X_1, \ldots, X_k$ satisfy the feasibility bounds discussed above, using the $\mathsf{pinf}$ value maintained by the executable solver. In this sense, they make up an approximate solution to the user-supplied problem.

\subsection*{Claude's Capability}

To illustrate Claude's skill at engineering Verus proofs, we outline how it proved a converse to soundness: a lifting lemma expressing that canonization preserves feasibility. Although this lemma is less important than soundness for gaining confidence in our solver, the process by which Claude proved it displays its ability to dynamically repair broken proofs.

Consider the predicate \lstinline[language=Verus]|user_feasible|, defined to hold on a list $\tau \coloneqq \mleft(X_1, \ldots, X_k\mright)$ of block decision variables if they are feasible (on the nose) for the user-supplied problem. The lifting lemma states that, as one would hope, if $\tau$ is user-feasible, then its lift is a feasible standard-form point, with or without the rank check:
\begin{lstlisting}[style=block,language=Verus]
proof fn lemma_user_feasible_lifts(
    sdp: &SDP, check: bool, s: &SdpadSetup,
    xs: Seq<Seq<real>>)
    requires
        sdpad_input_ok(sdp, check),
        sdpad_setup_ok(sdp, check, *s),
        sdp.lmi@.len() >= 1,
        user_feasible(sdp, xs),
    ensures ({
        let bd = bdims_r(sdp.lmi@);
        let vd = vdims_r(sdp.lmi@, n_vars(sdp));
        let z = lift_z(sdp.lmi@, xs, bd, vd);
        &&& is_psd(sym_view(z), s.var_dim as int)
        &&& ... })
\end{lstlisting}

Claude derives the ensured clause \lstinline[language=Verus]|is_psd(...)| from \cref{blockdec} combined with a separate lemma (not shown). The latter ensures that \lstinline[language=Verus]|lift_z|, which represents the lift of $\tau$, vanishes off the diagonal blocks and that each of its blocks is PSD. Both steps read individual entries of \lstinline[language=Verus]|lift_z|, an operation that depends on whether the given position is outside the blocks. That is, it involves the predicate expressing that $\mleft(r, c\mright)$ is a position lying above the diagonal and outside the blocks:
\begin{lstlisting}[style=block,language=Verus]
spec fn in_zero_region(bs: Seq<int>, r: int, c: int)
    -> bool {
    exists|j: int| 0 <= j < bs.len()
        && #[trigger] sum_upto(bs, j) <= r
               < sum_upto(bs, j + 1) <= c }
\end{lstlisting}
where \lstinline[language=Verus]|sum_upto(bs, j)| is the sum of the first $j$ block sizes, i.e., the offset at which block $j$ begins.

Reading an entry inside a block requires knowing that this predicate is false, i.e., a negated existential. But this formula becomes a universal over $j$ instantiated on the trigger \lstinline[language=Verus]|sum_upto(bs, j)|. This scenario can easily cause a \emph{matching loop}: an unbounded sequence of instantiations that exceeds the SMT solver's resource limit. Still, Claude repeatedly found it useful to keep the trigger as written.

Claude showed its skill by taking advantage of this trigger while preventing the matching loop. Specifically, it made the function that computes the entries of \lstinline[language=Verus]|lift_z| opaque:
\begin{lstlisting}[style=block,language=Verus]
#[verifier::opaque]
spec fn lift_entry(..., bd: Seq<int>, vd: Seq<int>,
    r: int, c: int) -> real {
    let bs = block_sizes(bd, vd);
    if in_zero_region(bs, r, c) { 0.0real }
    else { ... } }
\end{lstlisting}
Rather than remove the matching loop, the opaque attribute here confines it. Claude uses only one lemma (not shown) that reveals \lstinline[language=Verus]|lift_entry|'s body. This lemma's conclusion is also a case split on \lstinline[language=Verus]|in_zero_region|, so Z3 handles it without looking inside the universal, thus avoiding the matching loop.

\section{Conclusion}

We gave a nontrivial example of using an autonomous agent to formally verify a human-engineered codebase, namely a Rust implementation of the Wen--Goldfarb--Yin first-order method for semidefinite programming. For all its flaws, it illustrated a feasible approach to verifying a self-contained, relatively performant SDP solver. To us, the most attractive future direction is to replace the Jacobi eigensolver with the QR algorithm~\cite[Sec.~8.3]{golubvanloan} as the latter, though more elaborate, is faster. Even so, the eigensolver we offered here can readily be used for other pieces of verified software that rely on spectral decomposition.

\bibliography{sdpverusrefs}

\end{document}